\documentclass[12pt,preprint]{aastex}
\usepackage{graphics}
\usepackage{epsfig}
\usepackage{multirow}

\shorttitle{SDSS-WISE quasars}
\shortauthors{Wu et al.}

\begin{document}

\title{SDSS quasars in the WISE preliminary data release and quasar candidate selection with optical/infrared colors}

\author{Xue-Bing Wu\altaffilmark{1}, Guoqiang Hao\altaffilmark{1}, Zhendong Jia\altaffilmark{1}, Yanxia Zhang\altaffilmark{2}, Nanbo Peng\altaffilmark{2}}
\altaffiltext{1}{Department of Astronomy, School of Physics, Peking University, Beijing 100871, China; wuxb@pku.edu.cn}
\altaffiltext{2}{National Astronomical Observatories, Chinese Academy of Sciences, Datun Road A20, Beijing 100012, China}

\begin{abstract}
We present a catalog of 37842 quasars in the Sloan Digital Sky Survey (SDSS) Data Release 7, 
which have counterparts within 6$''$ in the Wide-field Infrared Survey Explorer (WISE) Preliminary Data Release. The overall WISE detection rate of the SDSS quasars is 86.7\%, and it 
decreases to less than 50.0\% when the quasar magnitude is fainter than $i=20.5$.
We derive the median color-redshift relations based on this SDSS-WISE quasar sample and apply
them to estimate the photometric redshifts of the SDSS-WISE quasars.
We find that by adding the WISE W1- and W2-band data to the SDSS photometry we can 
increase the photometric redshift reliability, defined as the percentage of sources with the 
photometric and spectroscopic redshift difference less than 0.2, from 70.3\% to 77.2\%.
We also obtain the samples of WISE-detected normal and late-type stars with SDSS spectroscopy, and present a criterion
in the $z-W1$ versus $g-z$ color-color diagram, $z-W1>0.66(g-z)+2.01$,  to separate quasars from stars. With this criterion we can recover 98.6\% of 3089 radio-detected SDSS-WISE quasars with redshifts less than four and overcome the difficulty in selecting quasars with redshifts between 2.2 and 3 from SDSS photometric data alone. We also suggest 
another criterion involving the WISE color only, $W1-W2>0.57$,  to
efficiently separate
quasars with redshifts less than 3.2 from stars. 
In addition, we compile a catalog of 5614 SDSS quasars detected by both WISE and UKIDSS surveys and present their color-redshift relations in the optical and infrared bands. By using the SDSS $ugriz$, UKIDSS YJHK and WISE W1- and W2-band
photometric data, we can efficiently select quasar candidates and 
increase the photometric redshift reliability up to 87.0\%. We discuss the implications of our results on the future quasar surveys. An updated SDSS–WISE quasar catalog
consisting of 101,853 quasars with the recently released WISE all-sky data is also
provided.
 
\end{abstract}

\section{Introduction}
Since the discovery of quasars in 1960s \citep{schmidt63}, more and more quasars have 
been found in the last four decades.  More than 120,000 quasars have 
been discovered from the recent large
optical spectroscopic surveys, such as the Two-Degree Fields(2dF) survey \citep{boyle00}
and the Sloan Digital Sky Survey (SDSS)\citep{york00, schneider10}. 
Quasar candidates in these surveys were mainly selected by
optical colors. Because of the strong UV and optical emission,
quasars can be usually distinguished from
the stellar objects in the color-color and color-magnitude diagrams based on optical
photometry \citep{smith05,richards02,fan00}.
However, quasar selections based on the SDSS optical photometry alone become very inefficient
for identifying $\rm 2.2<z<3.0$ quasars since they have optical
colors similar to those of normal stars \citep{fan99,richards02,richards06,schneider07}.
In order to get a more complete sample of quasars, we have to think about other ways to find these SDSS missing 
quasars in the 'redshift desert' ($\rm 2.2<z<3.0$).

An important way  to identify
the missing quasars with  $\rm 2.2<z<3.0$ has been suggested by using the infrared K-band excess based on the UKIRT
(UK Infrared Telescope) Infrared Deep Sky Survey (UKIDSS) \citep{warren00,hewett06,maddox08}.\footnote{The UKIDSS project is defined in \cite{lawrence07}. UKIDSS uses the UKIRT Wide Field Camera (WFCAM; 
\cite{casali07} and a photometric system described in \cite{hewett06}. 
The pipeline processing and science archive are described in 
\cite{hambly08}.} Although the $z\sim$2.7 quasars have
optical colors similar to stars, they are usually more luminous in the infrared
K-band. In addition, combining the optical colors in SDSS with the infrared 
colors in UKIDSS, it should be more efficient to separate stars from  both
lower redshift quasars ($z<$3)\citep{chiu07} and higher redshift ones 
($z>$6)
\citep{hewett06} in the color-color diagrams. Recent studies by \cite{wujia10}
and \cite{wu11} have demonstrated that with the SDSS-UKIDSS photometric data we can
efficiently select quasar candidates at $z<4$ with the selection criterion in the Y-K versus $g-z$ color-color diagrams. The spectroscopic observations carried out
by \cite{wu10a,wu10b} and \cite{wu11} have confirmed the effectiveness of using 
SDSS-UKIDSS colors to discover the missing SDSS quasars with  $\rm 2.2<z<3.0$.

However, the sky coverage of UKIDSS is limited and the extragalactic survey  of UKIDSS/LAS 
 will finally cover about 4000 $deg^2$ of the sky. Therefore, for a large part of the sky we can not use
the UKIDSS data in identifying quasars. 
The same problem exists for using the Spitzer Infrared Array Camera (IRAC) mid-infrared photometric data \citep{fazio04}, which also cover only very limited sky area, though the IRAC colors have been suggested to efficiently select quasars and other AGNs, including optically obscured quasars \citep{lacy04, stern05}.
In the near-IR bands, the
Two-Micron All-Sky Survey (2MASS)\citep{skrutskie06}) made an all-sky survey at J,H,K bands, 
reaching a depth of $K_s$=16 for sources with high Galactic latitudes. The 2MASS data have been used
to select quasars with B-J/J-K colors \citep{barkhouse01}. However, 2MASS is too shallow for most
quasars with magnitude $i>17$. Only 13930 of the 105783 quasars in SDSS DR7 have secure 2MASS photometry, though
53564 of them have 2$\sigma$ detections in at least one 2MASS band \citep{schneider10}.

Recently, the preliminary data release of NASA's Wide-field Infrared Survey Explorer (WISE) became publicly available \citep{wright10}. WISE has mapped all the sky at  3.4, 4.6, 12, and 22 $\micron$ in 2010 with an angular resolution of 6.1, 6.4, 6.5 and 12.0 arcsec and 5$\sigma$ photometric sensitivity better than 0.08, 0.11, 1 and 6 mJy (corresponding to 16.5, 15.5, 11.2, and 7.9 Vega magnitudes) in these four bands. The WISE  preliminary data release includes the
positional and photometric data for over 257 million objects with signal-to-noise ratio (S/N) greater than 7 in
at least one band, covering 23600 $deg^2$ of the sky  with the ecliptic longitude at $27.8^o<\lambda<133.4^o$ and $201.9^o<\lambda<309.6^o$. The final data release is scheduled in the 
spring of 2012. Therefore, the WISE all-sky photometric data will be very useful in helping us to select quasar candidates for the future quasar surveys if some selection criteria can be obtained with the known quasars with WISE detections in the preliminary data release.  
In addition, since the WISE bands are similar to those of Spitzer IRAC, we expect that the WISE colors can be also
adopted to select the reddened and optical obscured quasars, as already demonstrated by using the IRAC colors \citep{lacy04, stern05}. Taking the advantage of the full sky coverage of WISE data, a more complete quasar sample should be obtained. Although SDSS, 2dF and the ongoing SDSS III/BOSS surveys \citep{eisenstein11}
have discovered almost 200,000 quasars, the currently available quasar samples are still
incomplete due to the differences in quasar selection criteria and magnitude limits adopted for different surveys.  
Future efforts are still needed to find more currently missing quasars at different redshifts, including the reddened and obscured quasars, and construct a larger, deeper and more complete sample of quasars.

The paper is organized as follows. In Section 2 we present the SDSS-WISE quasar catalog. In Section 3 we
obtain the color-redshift relations for this SDSS-WISE quasar sample and investigate how they can improve the photometric redshift estimations of quasars. In Section 4 we propose the quasar candidate selection criteria based on our SDSS-WISE samples of spectroscopically confirmed quasars and stars. After analyzing a
SDSS-UKIDSS-WISE quasar sample in Section 5, we give our summary and discussion in Section 6.

\section{The SDSS-WISE quasar catalog}
We cross-correlate the sources in the quasar catalog of SDSS DR7 \citep{schneider10}, which consists of
105,783 SDSS quasars, with the sources in the WISE preliminary data release \citep{wright10}, which covers
the sky area with ecliptic longitude of $27.8<\lambda<133.4$ or $201.9<\lambda<309.6$ and
presents photometric information for over 257 million objects. Because the angular resolution of WISE is 6.1, 6.4, 6.5 and 12.0 arcsec in the four bands respectively, we use
6 arcsec as the position offset for finding the WISE counterparts of SDSS quasars.
Using a larger offset would lead to significant increase of duplicate WISE sources around 
SDSS quasars and higher rate of false positives in matching the SDSS-WISE catalogs. 
Because five SDSS quasars have more than one WISE counterparts within a 6 arcsec offset, we carefully exclude these duplicated WISE sources.  
The small number of duplicated WISE sources within a 6 arcseconds offset to the positions of SDSS quasars also indicates that the rate of false positives in our catalog matching should be very low. 
By also excluding also another five quasars without the full
detections in SDSS $ugriz$  bands, we create a catalog of 37842 SDSS quasars with WISE
detection in at least one of four WISE bands. 

In Figure 1 we show the histograms of the SDSS and WISE magnitudes, magnitude uncertainties, redshifts and 
position offsets of these SDSS-WISE quasars. The median value of each
quantity is also indicated in these histograms. The redshift range of these 37842 SDSS-WISE
quasars is from 0.064 to 5.414, with a median value of 1.442, and the $i$-band magnitude range is from 14.793 to 21.855, with a median value of 18.883.  
All the SDSS $ugriz$ magnitudes have been corrected from the Galactic extinction using
a map from \citep{schlegel98}. Throughout the paper the SDSS magnitudes are given in AB magnitudes, while
the WISE magnitudes are given in Vega magnitudes. The significant decreases of quasar numbers
in the histograms of W3- and W4-band magnitude uncertainties are due to the relatively lower
sensitivities of the WISE W3- and W4-band detectors. The median values of uncertainties of WISE W3 and W4 magnitudes are 0.127 and 0.283 respectively, which are
significantly larger than the median values 0.050 and 0.053 in the WISE W1 and W2 bands and the median values ($<0.043$) of the magnitude uncertainties in the SDSS $ugriz$ bands.
Although we use 6 arcsec for
the cross-correlation radius between the SDSS and WISE sources, from the position offset distributions
we see that majority of sources actually have offsets smaller than 2 arcsec. Therefore, we are
confident about the reliability of such an SDSS-WISE quasar catalog.

We also check the detection rate of SDSS quasars by WISE. There are 43662 sources in the
SDSS DR7 quasar catalog within the sky coverage of the WISE preliminary data release, so the overall
detection rate by WISE is about 86.7\%. In Figure 2 we show the redshift and $i$-band
magnitude histograms of both SDSS quasars in the sky area of the WISE preliminary data release
and the WISE detected SDSS quasars, as well as  the 
dependences of the WISE detection rate on
the redshift and magnitude.  From Figure 2 we can see that the WISE detection rate of SDSS
quasars is higher than 66\% at all redshift and is higher than 80\% at $z<2.2$, while it
is higher than 85\% at $i<19.5$  and decreases to lower than 50\% at  $i>20.5$. The lower
detection rate at  $i>20.5$ is understandable because of the limited sensitivity of WISE
detectors in the mid-infrared bands. 

In Table 1 we give the catalog of 37842 SDSS quasars detected in the WISE preliminary data 
release. The properties of these quasars, including the coordinates, offsets between the SDSS-WISE positions, redshifts, 
SDSS and WISE magnitudes and their uncertainties, and radio and X-ray properties (adopted from the SDSS DR7 quasar catalog of Schneider et al. 2010), are listed.   

Using the recently released WISE all-sky data we have compiled a new SDSS-WISE quasar catalog consisting of 101853 quasars, which is available in the online version of  Table 5. 

\section{Color-redshift relations and photometric redshift estimations}
With the SDSS-WISE quasar catalog, we can investigate their color-redshift relations, which are helpful to understand the quasar properties and can be used to
estimate the photometric redshifts of quasars.

From SDSS $ugriz$ and WISE W1,W2,W3,W4 magnitudes we can obtain eight colors for quasars. In Figure 3 we plot the color versus redshift diagrams for all the SDSS-WISE quasars. To obtain the
reliable color-redshift relations, we derive the median color-redshift relations
based on the quasars with magnitude uncertainties smaller than 0.2mag in $ugriz$ and the
W1,W2 bands and smaller than 0.4mag in the W3,W4 bands (corresponding to the
black dots in Figure 3).  
The SDSS quasars without detections in the WISE W3 and W4 bands are not included when calculating the colors related to these two bands. 
Clearly we fail to obtain
reliable $u-g$ color at $z>3.4$ and $g-r$ color at  $z>4.5$  because of the larger uncertainties of $u$ and $g$ magnitudes at larger redshifts as the quasar Ly$\alpha$ emission line moves out of the $u$ and $g$ filter bands. The larger
magnitude uncertainties in the WISE W3 and W4 bands also lead to  substantial
scatters in the color-redshift relations related to these magnitudes. To obtain
the reliable color-redshift relations, we only focus on quasars with $z<5$ because there are no enough quasars at $z>5$ in our SDSS-WISE catalog. We adopt
the bin size of 0.05 for $z<3$ and 0.1 for $z>3$ in order to have enough sources in each redshift bin to derive the median color. In Table 2, we give the median SDSS and WISE colors for quasars at redshifts from 0.075 to 5.

We need to keep in mind that using both SDSS and WISE data introduce selection bias to the SDSS-WISE quasar sample and also 
bias to the color-redshift relationships, since it is biased toward quasars that are intrinsically bluer and were selected by SDSS
as quasars. These biases are difficult to avoid when the currently largest quasar sample based on SDSS is adopted. In addition, our requirement for quasars detected in all SDSS  $ugriz$ bands and  using only quasars with smaller magnitude
uncertainties in constructing the median color-redshift relations can introduce bias too, which also make the median color
bluer. Although  we can treat the low S/N measurements with some statistical tools, we believe that using the relatively more accurate observational 
data is still the most direct and efficient way to derive reliable color-redshift relationships.
Therefore, our derived color-redshift relations based on the SDSS-WISE quasars may not be applicable to the reddened quasars and optically obscured 
Type 2 quasars. Reliable color-redshift relations of these special quasars will be obtained and compared with the current
results only
 when large samples of them are available in the future. Currently, it is still unclear what the fractions of
reddened quasars and Type 2 quasars are in the total quasar population \citep{richards03, glikman07,polleta08, reyes08}. Some techniques, involving mid-infrared and near-infrared data, have been proposed to find the
obscured quasars and reddened quasars \citep{lacy04, maddox08}. Similarly, we expect that the WISE data can also provide such helps in constructing a
more complete quasar sample.

With the derived color-redshift relations, we can use our previously established
$\chi^2$-minimization method to estimate the most probable photometric redshifts of quasars \citep{wu04, wujia10}. 
Here the  $\chi^2$ is defined as (see \cite{wu04}):
\begin{equation}
\chi^2=\sum_{ij}{\frac{[(m_{i,cz}-m_{j,cz})-(m_{i,observed}-m_{j,observed})]^2}
{\sigma_{m_{i, observed}}^2+\sigma_{m_{j,observed}}^2}},
\end{equation}
where the sum is obtained for all four SDSS colors and z-W1 and W1-W2 colors, $m_{i,cz}-m_{j,cz}$ is 
the color in the 
color-redshift relations, $m_{i,observed}-m_{j,observed}$ is the observed color 
of a quasar, and $\sigma_{m_{i, observed}}$ and $\sigma_{m_{j,observed}}$ are the
uncertainties of observed magnitudes in two SDSS-WISE bands. We do not use
the colors related to WISE W3 and W4 magnitudes because their uncertainties are 
substantially larger and only two third of sources in our SDSS-WISE catalog have
available values for the uncertainties of W4 magnitudes. 

We note that in using the simple form (Equation (1)) to calculate the $\chi^2$ values we need to assume that the 
measurements of colors and magnitude uncertainties are roughly in Gaussian distribution and uncorrelated,
which may not be true in the real case. Although the measurement of individual flux does follow a Gaussian
distribution, the measurements of magnitude and color generally do not follow it. Note that the SDSS magnitude
is asinh magnitude \citep{lupton99}, which introduces an extra complexity. Future efforts will be
needed to improve the $\chi^2$ calculation by considering the exact distributions of colors and magnitude uncertainties, 
though the result may not  change significantly. On the other hand, 
the colors of quasars appear to be
less correlated when comparing with stars, which can be easily observed from the color-color diagrams 
in the optical and near-infrared bands \citep{richards02, chiu07, maddox08}. The uncertainties
of magnitudes have also been shown to be minimally correlated \citep{weinstein04}. Therefore, the assumptions of non-correlations
of colors and magnitude uncertainties are believed to be reasonable. \cite{richards01} also adopted a similar formulae
as in Equation (1) to calculate the $\chi^2$ values (see their Equation (1)) by considering the constant scatters of the median
color-redshift relations. 

As in  \cite{wujia10},
in order to compare the $\chi^2$ values for
the cases where different number of colors at different redshifts were used, we
actually use the $\chi^2/N$ (where N is the number of colors and is 6,5 and 4
respectively for the input redshift of $z<3.4$,$3.4<z<4.5$ and $z>4.5$) instead of 
$\chi^2$ to determine the photometric redshift by obtaining the minimum of
$\chi^2/N$ at a certain redshift. An IDL program is made to search the
photometric redshifts of SDSS-WISE quasars by taking the above factors into account.

In order to see more clearly whether using the WISE colors can 
improve the photometric redshift estimation, we also estimate the photometric redshifts
of these quasars using the SDSS colors only. In Figure 4 we compare the results obtained
by SDSS and by SDSS+WISE colors. We can see that by adding WISE W1 and W2 magnitudes we can improve
the photometric redshift estimations substantially. If we use the SDSS colors alone, the 
photometric redshift reliability, defined as the percentage of sources with the 
photometric and spectroscopic redshift difference ($|z_{photo}-z_{spec}|$) less than 0.2, is 70.3\%. If we
add z-W1 and W1-W2 colors to SDSS colors, such reliability increases to 77.2\%.
Especially for SDSS-WISE quasars with $i<19.1$ and $i<20.5$,  which correspond roughly to the depth of the SDSS and WISE quasar samples respectively, using the SDSS colors alone leads to the 
photometric redshifts reliability of 72.82\%  and 70.46\%,  while  adding the WISE colors can increase 
the reliability to 79.97\% and 77.48\%, respectively. For quasars with redshifts between 2.2
and 3, which have optical colors similar to stars and are difficult to  select by optical colors,  the photometric redshift reliability
can increase from 67.89\% (62.64\%)  using when SDSS colors alone to 75.25\% (69.41\%)  using when SDSS+WISE
colors if they are brighter than $i=19.1$(20.5). 
This clearly demonstrates the effectiveness of adding the WISE infrared colors in the
 photometric redshift estimations for quasar samples with different magnitude limits and redshift ranges. 

\section{Quasar candidate selections with SDSS and WISE photometric data}

One of the most important things for optical quasar surveys is to efficiently
select the quasar candidates. In SDSS, quasar candidates are mostly selected based on
the multi-band optical photometric data \citep{richards02}. However, SDSS quasar
selection is very inefficient at redshifts between 2.2 and 3 due to the similar 
optical colors of quasars with redshifts in this range to those of stars \citep{warren00}. One possible
way to improve this situation is to use the near-IR colors \citep{warren00,
hewett06, maddox08}. Because quasars usually have a much flatter spectral energy
distribution over a wide range of wavelengths, their spectral shapes in the near-IR bands are
different from those of normal stars even if their optical spectra are similar to stars (e.g. for quasars
with redshifts between 2.2 and 3). \cite{wujia10}
have demonstrated that by combining the UKIDSS near-IR colors with SDSS optical colors we can 
separate well quasars from stars, and efficiently select quasars with redshift less than five. However, because UKIDSS/LAS will only cover 
the sky area of 4000 $deg^2$, we have to think about other ways to improve the quasar
selection method. Here we investigate the cases of using the data in the WISE W1 and W2 bands, which
are close to the near-IR bands. 

\cite{wujia10}  have suggested using $Y-K>0.46(g-z)+0.82$ (here g and z are AB magnitudes and Y and K are Vega magnitudes, see also \cite{wu11}) to efficiently separate quasars with redshifts
$z<4$ from stars in the Y-K versus $g-z$
color-color diagram. We think that this may still be the case in the $z$-W1 versus $g-z$
diagram because $z$ and W1 bands are close to Y and K bands respectively. In order to check this idea, we obtain a sample of normal stars and a
sample of late-type stars with both
SDSS DR7 spectroscopy and WISE data. Similar to what we did for SDSS-WISE quasars, we adopted
6$''$ as the offset between the SDSS and WISE positions for the star 
and late-type star samples, and deleted a few duplicated WISE sources with relatively
larger offsets. In order to get a reliable quasar selection criterion, we also
include only the quasars and stars with magnitude uncertainties in the $g$, $z$ and W1 bands 
less than 0.2 mag. Finally we adopt the SDSS-WISE samples of 37,535 quasars, 19,765
normal stars and 15,359 late-type stars to investigate the criterion of
separating quasars from stars. We note that the SDSS spectroscopically identified star sample
is very biased as many of them have similar optical colors to quasars, but we believe that 
including more stars with very different optical colors from quasars will not affect our
quasar selection criterion because these stars should be well separated from quasars in our color-color diagram
than the stars with optical colors similar to quasars.

In Figure 5 we plot the distributions of these SDSS-WISE quasars and stars in the 
z-W1 versus $g-z$ diagram. Obviously, most quasars with redshifts less than four can be 
separated from both normal and late-type stars on this color-color diagram. 
This is also confirmed by the median z-W1 and $g-z$ colors at different redshift,
shown as yellow solid line in Figure 5, which is obtained from the median color-redshift
relation of SDSS-WISE quasars. However,
there are significant overlaps, especially between quasars with $z>4$ and stars.
Similar to in \cite{wujia10}, we perform an automatic search for the best criterion 
to efficiently separate quasars and stars. We obtain this
criterion as:  $z-W1=0.66(g-z)+2.01$. With this criterion, we can select 36895 of
37,535 quasars (with a percentage of 98.30\%) and select 33,442 of 35,124 stars (with a percentage of 95.21\%). The false positive rate, defined as the ratio between the number 
of stars (1682) incorrectly selected as quasars and the number of all sources selected as
quasars (38,577) by our criterion, is 4.36\%. However, we must keep in mind that the actual number of
stars could be significantly larger than what we used here because only a tiny fraction of stars have been 
spectroscopically observed by SDSS. Therefore, the real false positive rate of our quasar selection may be higher.
For 37,272 SDSS-WISE quasars with redshifts less than four, with the proposed criterion
we can select 36,827 of them, with  completeness of 98.81\%.
This demonstrates the very high efficiency
in selecting  $z<4$ quasars with the $z-W1/g-z$ criterion, similar to using the
 $Y-K/g-z$ criterion \citep{wujia10}.   
 
We also explore the case where we use WISE colors only to select quasars. 
For this purpose, we selected 37,816 quasars, 19,369 normal stars and 18,127 late-type stars with both SDSS DR7 spectroscopy
and WISE W1,W2 and W3 band data. 
In the upper panel of 
Figure 6 we show the distributions
of SDSS-WISE quasars and stars in the W1-W2 versus W2-W3 diagram. In the lower panel of Figure 6 we show the histograms of W1-W2 colors of stars and quasars with different redshifts. 
Clearly, quasars with redshifts smaller than
3.5 are separated from stars by the W1-W2 color, but quasars with higher redshifts largely overlap with
stars. 
Such distributions are similar as that found by the Spitzer IRAC colors \citep{lacy04, stern05}.
We also did a search for the best criterion to separate the quasars and stars
and obtained this criterion as $W1-W2>0.57$ (shown as the dashed line in Figure 6).
Using this criterion, we can select 36,565 from 37,816 quasars, 18,837 from 19,369 normal stars, and 17,922 from
18,127 late-type stars. 96.69\% of the SDSS-WISE quasars and 98.04\% of SDSS-WISE stars can be separated
with this criterion. The false positive rate, due to the incorrect classification of 737 stars as quasars
by this W1-W2 criterion, is 1.98\%. Comparing with the case of using the previous criterion in the
z-W1/g-z diagram, using the W1-W2 criterion can lead to lower false positive rate and also lower 
completeness of selecting quasars. 

In order to check whether using our proposed two quasar selection criteria can avoid the color bias
in the SDSS quasar selection algorithm, we obtain a sample of 3089 FIRST radio selected SDSS-WISE quasars.
Because these radio-detected quasars are spectroscopically identified in SDSS without involving
the optical/infrared color selections, they are often adopted to check the quasar selection efficiency in using the proposed
selection criteria \citep{richards06}. In Figure 7 we demonstrate the results of using two quasar selection criteria.
From Figure 7 we can clearly see that with our $z-W1/g-z$ selection criterion, we can recover 3035 of the 3089
radio detected quasars  at a completeness of 98.25\%. This completeness raises to 98.63\% for radio
quasars with $z<4$, which confirms the robustness of the  $z-W1/g-z$ selection criterion.
With the W1-W2 selection criterion, we can recover 2989 of the 3089
radio detected quasars  at a completeness of 96.76\% and such completeness raises to 97.97\% for radio
quasars with $z<3.2$ (note that the 'noise' at $z>3.5$ in Figure 7 is due to the small number statistics). 
Our investigations demonstrate that the W1-W2 criterion and  the $z-W1/g-z$ 
 criterion can be adopted to efficiently select  $z<3.2$ and  $z<4$ quasars respectively with very high
completeness. However, to selecte high redshift quasars with $z>4$, obviously we still 
need to use other selection criteria (see \cite{fan01, hewett06, wujia10}).

To better understand whether our proposed quasar selection criteria are efficient
in selecting quasars down to the magnitude limits of $i=19.1$ and $i=20.5$, which correspond roughly to the depth of the SDSS and WISE quasar samples respectively, we made the following checks with our SDSS-WISE quasar sample. For 37842 quasars in this sample, there are 26397 quasars with magnitudes brighter than $i=19.1$. Using our proposed  $z-W1/g-z$  and W1-W2 selection criteria we can recover 26091 and 25977 of them at a completeness of 98.84\% and 98.41\%, respectively. For 1927 quasars with magnitudes brighter than $i=19.1$ and redshifts between 2.2 and 3, we can recover 97.15\% and 97.46\% of them with these two criteria. For 37609 quasars with magnitudes brighter than $i=20.5$ in our SDSS-WISE sample, using the $z-W1/g-z$  and W1-W2 criteria we can recover 36863 and 36373 of them at a completeness of 98.02\% and 96.71\%, respectively. For  3043 quasars with magnitudes brighter than $i=20.5$ and redshifts between 2.2 and 3, we can recover 96.94\% and 95.40\% of them with these two criteria. Obviously these checks demonstrate that with our proposed quasar selection criteria we can efficiently select WISE detected quasars even at the magnitude limit down to $i=20.5$. Specifically, our criteria can be used to recover the quasars with
redshifts between 2.2 and 3 very efficiently, which can be also seen from Figure 7.  This may have
important implications on the quasar candidate selections for future spectroscopic quasar surveys.

\section{The SDSS-UKIDSS-WISE quasars}
With the SDSS-WISE quasar sample, we can also find the UKIDSS counterparts for some of these quasars. In this case,
we are able to construct a quasar sample with the photometric data from SDSS, UKIDSS to WISE bands. Using
the DR6 public data of the UKIDSS/LAS, we obtain this SDSS-UKIDSS-WISE quasar sample, which consists of 5614 quasars
with offsets between the SDSS and UKIDSS positions within 3 arcsec and with all detections at
SDSS $ugriz$, UKIDSS YJHK and WISE W1 and W2 bands. We do not include the WISE W3 and W4 data because of the
relatively lower sensitivities in these two bands. Requiring the detections in W3 and W4 bands will
substantially reduce the quasar number of our sample. The data of these SDSS-UKIDSS-WISE quasars, including the coordinates, redshifts
and 11-band magnitudes ($ugriz$ in AB magnitudes, YJHK and W1,W2 in Vega magnitudes), are given in Table 3.
From the photometric data in 11 bands, we can construct
the color-redshift relations of these SDSS-UKIDSS-WISE quasars, which are shown in Figure 8. 
The median relations are also obtained from the data
with magnitude uncertainties at all bands less than 0.2mag, and are summarized in Table 4. 

With the color-redshift relations of SDSS-UKIDSS-WISE quasars, we can obtain
the photometric redshifts with different sets of photometric data using the
techniques described in Section 3, and compare
the photometric redshift reliability obtained with the different photometric 
data. In Figure 9, we show the comparisons of photometric redshifts with 
spectroscopic redshifts
and the distributions of the differences between them for the cases 
using the SDSS, SDSS+UKIDSS, UKIDSS+WISE W1,W2, and SDSS+UKIDSS+WISE W1,W2 photometric data, respectively. The photometric redshift reliability, defined as the
fraction of the sources with the difference between the photometric and spectroscopic redshifts smaller than
0.2, is 70.4\%, 84.8\%, 67.4\% and 87.0\% respectively for the above four cases.  Therefore, by using the SDSS $ugriz,$ UKIDSS YJHK and WISE W1 and W2 band
photometric data, we can efficiently 
improve the photometric redshift reliability up to 87.0\%, which is significantly higher than using the SDSS
photometric data alone (70.4\%). Moreover, as mentioned in Section 4 and in \cite{wujia10}, using  the SDSS,
UKIDSS and WISE photometric data  can also help us  select quasar candidates more efficiently. However,
we noticed that this can be done only in the UKIDSS surveyed area, which is much smaller than the sky
coverage of both SDSS and WISE surveys.

\section{Summary and Discussion}

In this paper, we present a catalog of 37842 SDSS quasars having counterparts  in the WISE Preliminary Data Release within 6$''$. The overall WISE detection rate of the SDSS quasars in the sky area of the WISE preliminary data release is 86.7\%, which demonstrates that the WISE data can be very helpful in identifying
quasars, especially  those with magnitudes brighter than $i=20.5$.
By deriving the median color-redshift relations of this SDSS-WISE quasar sample, 
we develop a method to estimate   the photometric redshifts of quasars and find that  the photometric redshift reliability can 
increase from 70.3\% to 77.2\% if the WISE W1- and W2-band data are added to the SDSS photometry.
We also obtain a criterion
in the z-W1 versus g-z color-color diagram, $z-W1>0.66(g-z)+2.01$,  to separate quasars from stars. With this criterion we can recover 98.6\% of 3089 radio-detected SDSS-WISE quasars with redshifts less than four and overcome the difficulty in selecting quasars with redshifts between 2.2 and 3 from the SDSS photometric data alone. We also suggest 
another criterion involving the WISE color only, $W1-W2>0.57$, to separate
quasars with redshifts less than 3.2 from stars. 
In addition, we compile a catalog of 5614 SDSS quasars detected by both WISE and UKIDSS surveys and present their color-redshift relations. By using the SDSS $ugriz$, UKIDSS YJHK and WISE W1- and W2-band
photometric data, we can efficiently select quasar candidates and 
increase the photometric redshift reliability up to 87.0\%. 

Considering the advantages of all-sky coverage of  the WISE mid-infrared photometry,  the WISE data will
be very helpful in finding new quasars in  future quasar survey and constructing a more complete quasar
sample than that currently available.   The ongoing BOSS project in SDSS III has identified 29,000 quasars with $z>2.2$ and expects to obtain
the spectra of 150,000 quasars at  $2.2<z<4$ \citep{eisenstein11}, with the updated quasar target selection techniques including K-band excess \citep{ross12}.  The WISE data may be also helpful to the BOSS quasar target selection, especially in selecting quasars with $i<20.5$. The Chinese GuoShouJing telescope (LAMOST)\citep{su98}, which is a 4-meter size spectroscopic telescope with 4000 fibers and 5-degree field of view  and
is currently in the commissioning phase, 
is also aiming to discover 0.4 million quasars
with magnitudes bright than $i=20.5$ in the next four years \citep{wu10b}. The WISE data will be adopted
to select quasar candidates in the LAMOST quasar survey. Obviously,  the results obtained from this 
paper, especially the proposed quasar selection criteria and the photometric redshift estimation methods, will
provide significant help in selecting LAMOST quasar candidates in a large sky areas.

We have demonstrated the advantages of combining the SDSS optical and UKIDSS, WISE infrared photometric
data in finding quasars. However, in order to fully use this advantage to discover more quasars we still need much wider and deeper photometry in the
optical and infrared bands.
Fortunately, several ongoing and upcoming  photometric sky surveys will provide such
 helps to us.   SDSS III \citep{eisenstein11} has taken 2500 deg$^2$ 
further imaging in $ugriz$ bands
in the south galactic cap aside from the SDSS I and II photometry. The SkyMapper \citep{keller07} and Dark Energy Survey \citep{the05} will also present the multi-band optical 
photometry in 20000/5000 deg$^2$ of the southern sky, reaching the magnitude limit of 22/24 mag in the $i$-band,
respectively. The Visible and Infrared Survey Telescope for Astronomy (VISTA)\citep{arnaboldi07} will 
carry out its
VISTA Hemisphere Survey  in the near-IR YJHK bands for 20000 deg$^2$ of the southern sky with
a magnitude limit at K=20.0, which is about 5 mag and 2 mag deeper than the 2MASS
and UKIDSS limits, respectively. Therefore, the WISE data, when combined with the optical and infrared photometric data
obtained with these ongoing and upcoming surveys, will provide us with a large database for quasar candidate selections using the proposed optical/infrared selection criteria. 
 We expect that a much larger and more complete quasar sample covering a wider range of
 redshift  will be constructed in the near future, which will play an important role in the near future
in studying extragalactic astrophysics, including  AGN physics, galaxy  
evolution, large scale structure and cosmology.

\acknowledgments 

We thank the anonymous referee for  very helpful comments and suggestions.  The work is supported by the National Natural Science Foundation of China (11033001) and the National Key Basic Research Science Foundation of China (2007CB815405).

This publication makes use of data products from the Wide-field Infrared Survey Explorer, which is a joint project of the University of California, Los Angeles, and the Jet Propulsion Laboratory/California Institute of Technology, funded by the National Aeronautics and Space Administration.

Funding for the SDSS and SDSS-II has been provided by the Alfred P. Sloan Foundation, the Participating Institutions, the National Science Foundation, the US Department of Energy, the National Aeronautics and Space Administration, the Japanese Monbukagakusho, the Max Planck Society and the Higher Education Funding Council for England. The SDSS web site is http://www.sdss.org/.

The SDSS is managed by the Astrophysical Research Consortium for the Participating Institutions. The Participating Institutions are the American Museum of Natural History, Astrophysical Institute Potsdam, University of Basel, University of Cambridge, Case Western Reserve University, University of Chicago, Drexel University, Fermilab, the Institute for Advanced Study, the Japan Participation Group, Johns Hopkins University, the Joint Institute for Nuclear Astrophysics, the Kavli Institute for Particle Astrophysics and Cosmology, the Korean Scientist Group, the Chinese Academy of Sciences (LAMOST), Los Alamos National Laboratory, the Max-Planck-Institute for Astronomy (MPIA), the Max-Planck-Institute for Astrophysics (MPA), New Mexico State University, Ohio State University, University of Pittsburgh, University of Portsmouth, Princeton University, the United States Naval Observatory and the University of Washington.

{\it Facilities:} \facility{Sloan (SDSS)}, \facility{UKIRT}, \facility{WISE}

\newpage

\begin{table}
\setlength{\tabcolsep}{4pt}
{\scriptsize \caption{A catalog of 37,842 SDSS-WISE quasars}
\begin{tabular}{ccccccccccccc}
\hline \noalign{\smallskip}
RA & Dec & offset & redshift & u & g & r & i & z & W1 & W2 & W3 & W4\\
deg& deg  & $''$   &   &  &  &  & & & & & \\
\noalign{\smallskip} \hline \noalign{\smallskip}
   18.984833  &  31.799980  &  0.853  &  1.265  &   20.158   &  19.792  &   19.192  &   19.027   &  18.944   &    15.163   &  13.754 &10.814  &   8.511  \\   
   19.005892  &   32.376339  &  0.141  &  0.662  &   21.666  &   20.586  &   20.334  &   20.145   &  20.031   &    15.177   &  14.193 &11.805 &    8.927 \\   
   19.012709  &   31.423994  &  1.979  &  2.645   &  20.677  &   19.831  &   19.630  &   19.550  &   19.345  &    15.690  &   15.013 &11.940  &   8.370   \\  
   19.063026   &  32.578110   & 0.859  &  2.389  &   21.790   &  20.524   &  20.390   &  20.116   &  19.824   &   16.458   &  15.487& 11.813  &   9.065 \\   
   19.193468   &  31.414030  &  2.501   & 2.005   &  21.009   &  20.409   &  20.071   &  19.769  &   19.424   &    15.665   &  14.573&11.804  &   8.836  \\ 
\noalign{\smallskip} \hline \noalign{\smallskip}
\end{tabular}\\
}
{\scriptsize Notes: This table is available in its entirety in machine-readable and Virtual
Observatory (VO) forms in the online journal. A portion is shown here for guidance regarding its form and content. } 
\end{table}

\begin{table}
{\scriptsize \caption{The color-redshift relations of SDSS-WISE quasars}
\begin{tabular}{ccccccccc}
\hline \noalign{\smallskip}
Redshift & u-g & g-r & r-i & i-z & z-W1 & W1-W2 & W2-W3 & W3-W4\\
\noalign{\smallskip} \hline \noalign{\smallskip}
  0.075 &  0.259  &  0.097&   0.297 & -0.141 &  4.387 &  0.963 &  2.860 &  2.406\\
  0.125  &  0.037 &  0.112&   0.475 & -0.140 &  4.343 &  0.962 &  2.763 &  2.281\\
  0.175  &  0.049 &  0.205&   0.387 & -0.043 &  4.349 &  0.943 &  2.743 &  2.356\\
  0.225  &  0.093 &  0.278&   0.323 &  0.014 &  4.242 &  0.957 &  2.753 &  2.321\\
  0.275 &   0.089 &  0.257 &  0.085 &  0.410 &  4.023 &  0.962 &  2.766&   2.367\\
\noalign{\smallskip} \hline \noalign{\smallskip}
\end{tabular}\\
}
{\scriptsize Notes: This table is available in its entirety in machine-readable and Virtual
Observatory (VO) forms in the online journal. A portion is shown here for guidance regarding its form and content. } 
\end{table}

\begin{table}
\setlength{\tabcolsep}{3pt}
{\scriptsize \caption{A catalog of 5614 SDSS-UKIDSS-WISE quasars}
\begin{tabular}{cccccccccccccc}
\hline \noalign{\smallskip}
RA & Dec  & redshift & u & g & r & i & z & Y & J & H & K & W1 & W2\\
deg& deg  &    &   &  &  &  & & & & & & &\\
\noalign{\smallskip} \hline \noalign{\smallskip}

   25.994659  &   14.706349 &    1.634 &  18.984 &  18.821 &  18.749 &  18.634 &  18.608&   18.389 &  17.868 &  17.413 &  17.148 &  15.907 &  14.114\\   
   26.124020  &   14.556283  &    1.615  & 19.385  & 19.237  & 19.133 &  18.863  & 18.906  & 18.568  & 18.168  & 17.429   &16.943  & 16.043  & 14.848 \\ 
   26.393171   &  14.526940  &    0.635  & 20.106  & 19.670  & 19.444 &  19.131  & 19.037 &  18.361 &  18.046  & 17.198  & 16.063  & 14.442  & 13.328 \\  
   26.403246    & 14.928178   &   1.147 &  17.873 &  17.702 &  17.539  & 17.535 &  17.584  & 16.944  & 16.932  & 16.488  & 15.802  & 14.386  & 12.932 \\  
   26.595579   &  14.804535   &    0.971 &  18.787  & 18.757 &  18.547  & 18.729   &18.706  & 18.677 &  17.815  & 17.631 &  16.605   &14.966 &  13.755 \\  
\noalign{\smallskip} \hline \noalign{\smallskip}
\end{tabular}\\
}
{\scriptsize Notes: This table is available in its entirety in machine-readable and Virtual
Observatory (VO) forms in the online journal. A portion is shown here for guidance regarding its form and content.} 
\end{table}

\begin{table}
{\scriptsize \caption{The color-redshift relations of SDSS-UKIDSS-WISE quasars}
\begin{tabular}{ccccccccccc}
\hline \noalign{\smallskip}
Redshift & u-g & g-r & r-i & i-z & z-Y & Y-J & J-H & H-K & K-W1 & W1-W2\\
\noalign{\smallskip} \hline \noalign{\smallskip}
 0.075 &  0.114 &  0.086 &  0.297 & -0.157 &  0.396 &  0.395 &  0.304 &  0.716  &  1.085 &  0.963\\
  0.125 &  0.087  & 0.115 &  0.442  &-0.209 &  0.498 &  0.602 &  0.736 &  1.117  &  1.334  & 0.996\\
  0.175 &  0.024 &  0.178 &  0.391 & -0.070 &  0.759 &  0.607 &  0.670 &  0.963  &  1.347  & 0.963\\
  0.225 &  0.073 &  0.230 &  0.346 & -0.016 &  0.714 &  0.522 &  0.778 &  0.961  &  1.275 &  0.964\\
  0.275  & 0.111 &  0.276 &  0.086 &  0.414  & 0.574 &  0.516  & 0.800 &  0.961 &   1.234 &  0.959\\
\noalign{\smallskip} \hline \noalign{\smallskip}
\end{tabular}\\
}
{\scriptsize Notes: This table is available in its entirety in machine-readable and Virtual
Observatory (VO) forms in the online journal. A portion is shown here for guidance regarding its form and content.} 
\end{table}

\begin{table}
\setlength{\tabcolsep}{4pt}
{\scriptsize \caption{A catalog of 101,853 SDSS-WISE quasars by atching the WISE all-sky data}
\begin{tabular}{ccccccccccccc}
\hline \noalign{\smallskip}
RA & Dec & offset & redshift & u & g & r & i & z & W1 & W2 & W3 & W4\\
deg& deg  & $''$   &   &  &  &  & & & & & \\
\noalign{\smallskip} \hline \noalign{\smallskip}
 
0.033900& 0.276301& 0.387 &1.837 &20.242& 20.206 &19.941& 19.485& 19.178 &16.170& 14.711& 11.424 &8.139\\
 0.038604 &15.298476& 0.209 &1.199 &19.916& 19.807& 19.374& 19.148& 19.312& 15.656& 14.084& 10.935 &8.388\\ 
0.039089 &13.938449 &0.128 &2.234 &19.233& 18.886& 18.427 &18.301 &18.084 &15.513 &14.672& 10.825 &8.750\\ 
0.039271 &-10.464425& 0.465& 1.845& 19.242& 19.019& 18.966& 18.775& 18.705& 16.158 &15.099& 12.022 &8.865\\ 
0.047549 &14.929355 &0.180& 0.460& 19.647& 19.465& 19.368 &19.193& 19.015& 14.263& 13.200& 10.769 &8.158\\

\noalign{\smallskip} \hline \noalign{\smallskip}
\end{tabular}\\
}
{\scriptsize Notes: This table is available in its entirety in machine-readable and Virtual
Observatory (VO) forms in the online journal. A portion is shown here for guidance regarding its form and content. } 
\end{table}

\begin{figure}
\plotone{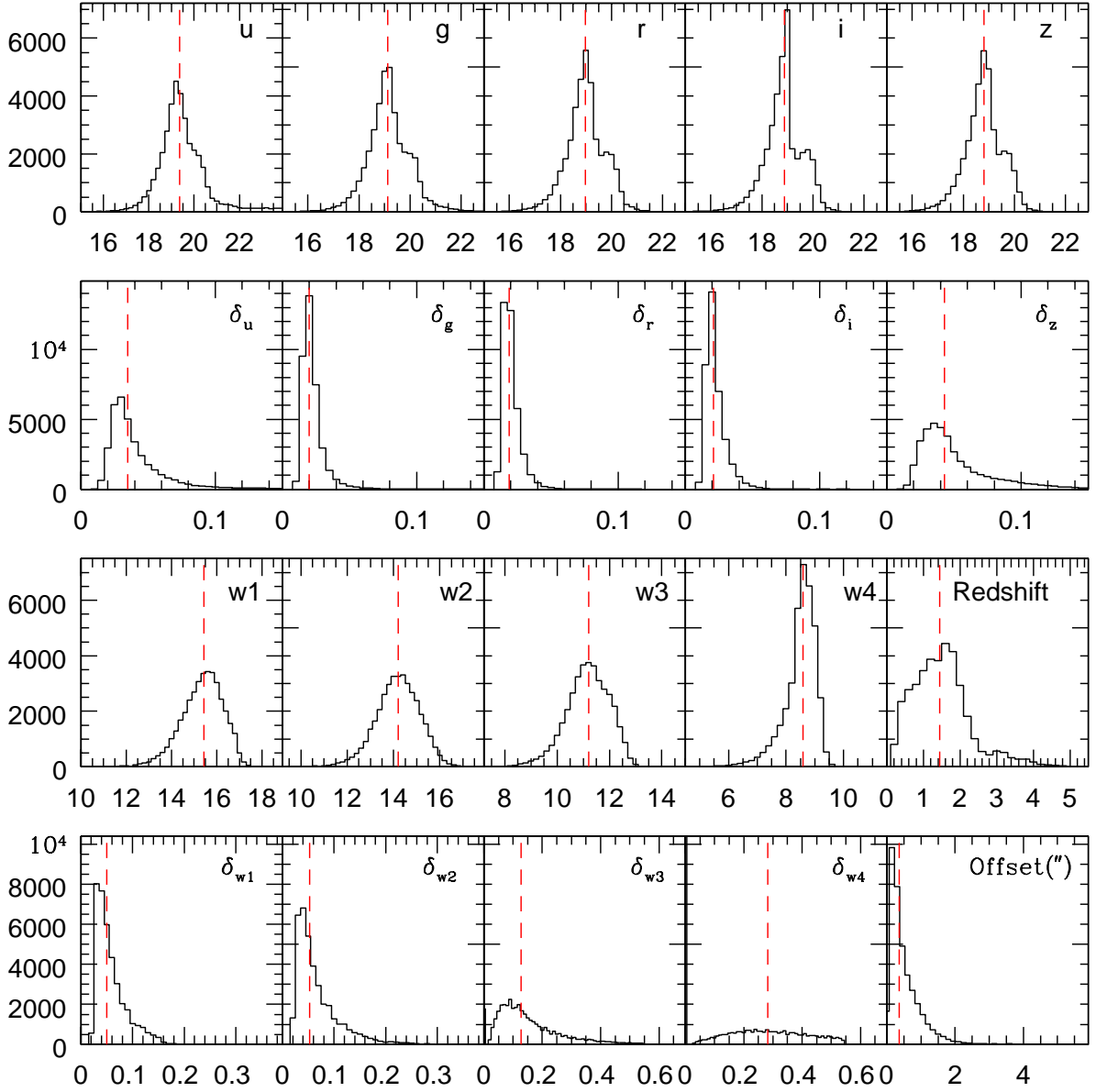}
\caption{Histograms of magnitudes, magnitude uncertainties, redshifts and offsets
between the SDSS and WISE positions of the SDSS-WISE quasars. The dashed line marks the median value
of each quantity. }
\end{figure}

\begin{figure}
\plotone{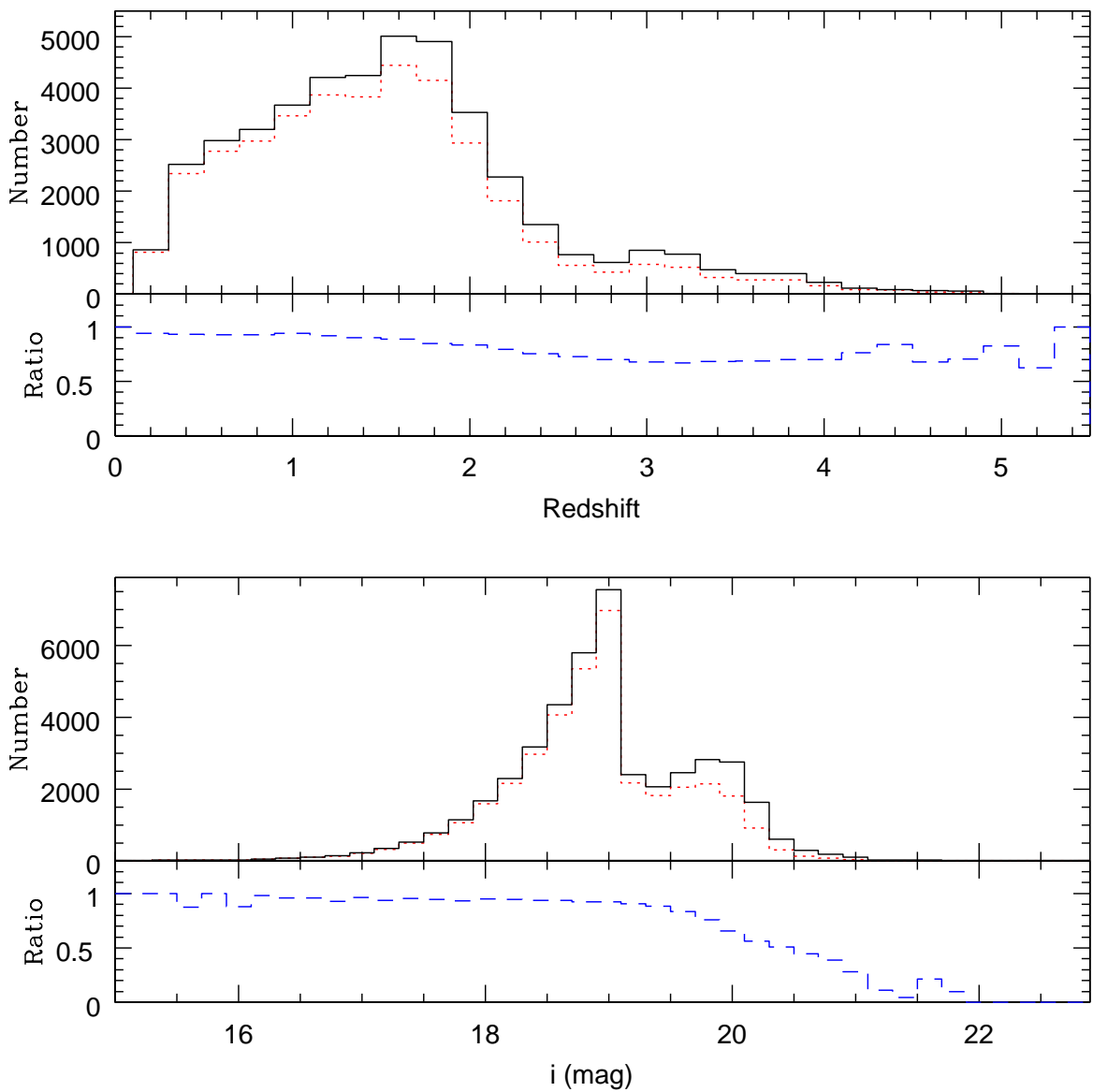}
\caption{Upper panel: The solid line denotes the redshift distribution of SDSS quasars in the sky area of WISE preliminary data release, while the dotted line denotes the redshift histogram
of the WISE detected SDSS quasars. The ratio between them is also plotted as a function of redshift. Lower panel: The solid line denotes the $i$-band magnitude distribution of SDSS quasars in the sky area of WISE preliminary data release, while the dotted line denotes the $i$-band magnitude histogram
of the WISE detected SDSS quasars. The ratio between them is also plotted as a function of magnitude. }
\end{figure}

\begin{figure}
\plotone{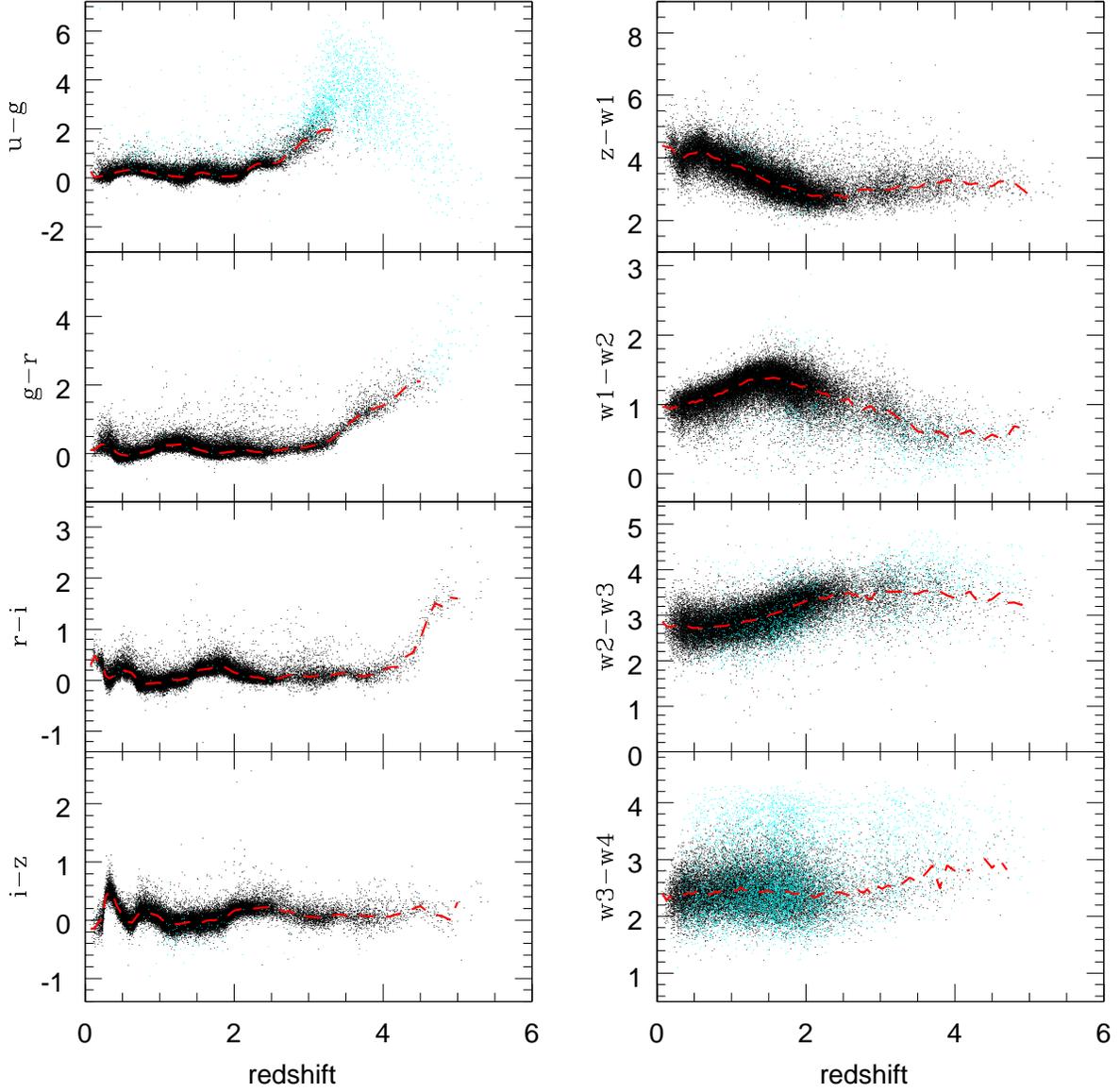}
\caption{Color-redshift relations of SDSS-WISE quasars. The black dots
denote the quasars with magnitude uncertainties smaller than 0.2mag in $ugriz$ and W1, W2 bands and smaller than 0.4 mag in W3, W4 bands.  Other quasars with
larger magnitude uncertainties are plotted as cyan dots. The dashed lines represent the median color-redshift relations obtained from the quasars denoted as
black dots.
}
\end{figure}

\begin{figure}
\plotone{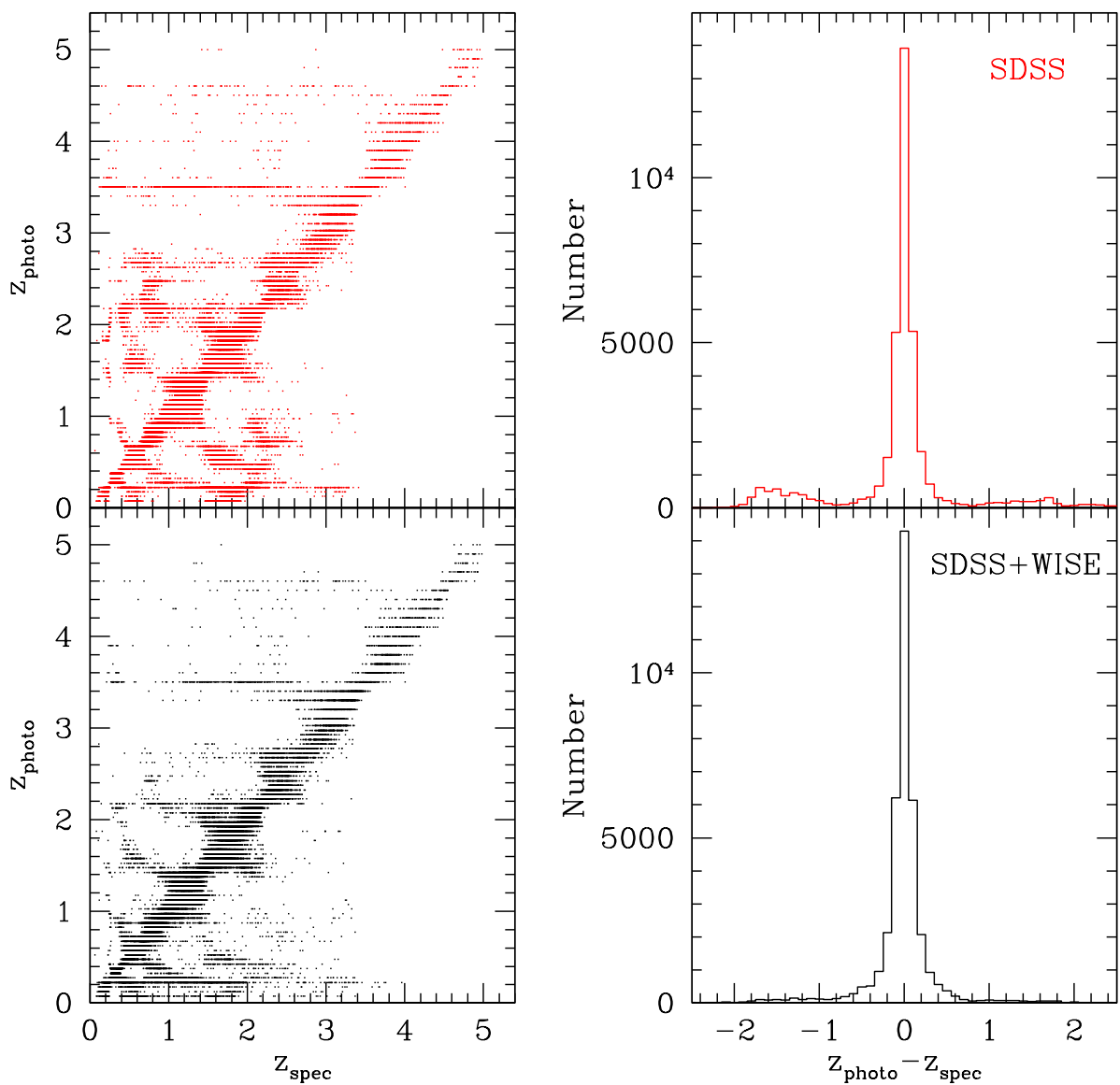}
\caption{Comparisons of photometric redshifts with spectroscopic redshifts
and the distributions of the differences between them. The upper panels show the cases for
using the SDSS $ugriz$ magnitudes and the lower panels show the cases for using
SDSS $ugriz$ and WISE W1,W2 magnitudes. The improvements in the later cases can be
clearly seen.
}
\end{figure}

\begin{figure}
\plotone{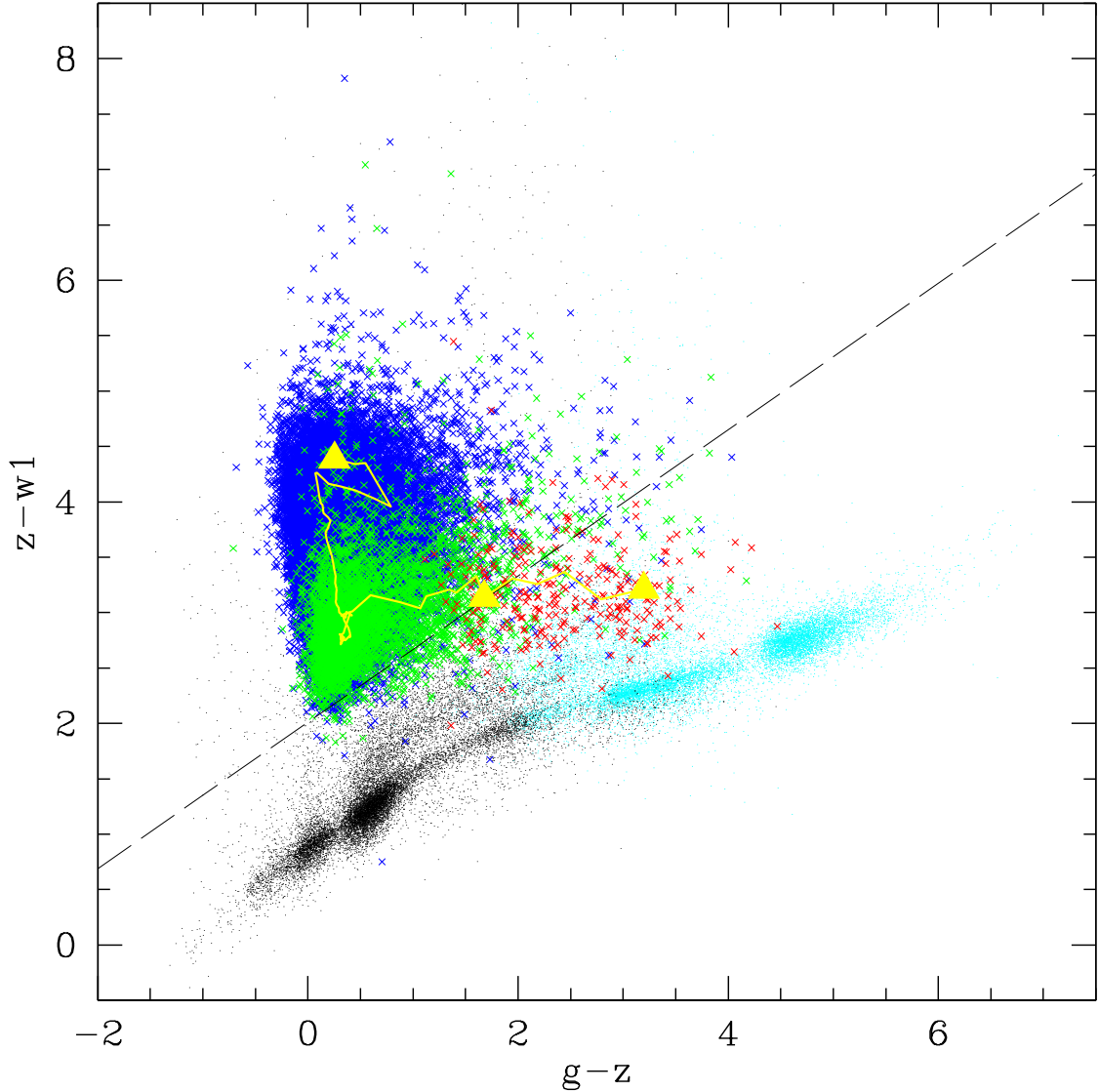}
\caption{Distributions of SDSS-WISE quasars and stars in the $z-W1$ vs. $g-z$ 
color-color diagram. The blue, green and red crosses denote the quasars with redshifts of
$z<2.2$, $2.2<z<4$ and $z>4$, respectively. The black and cyan dots denote the
SDSS identified normal stars and late-type stars. The yellow line represents the
median color-color relation derived from the median color-redshift relation
of SDSS-WISE quasars, and the yellow triangles, from left to right, mark the quasars with redshifts of
$z$=0.1, 4, and 4.5 in the median color-color relation. The dashed line indicates our proposed quasar selection criterion, $z-W1>0.66(g-z)+2.01$.
}
\end{figure}

\begin{figure}
\plotone{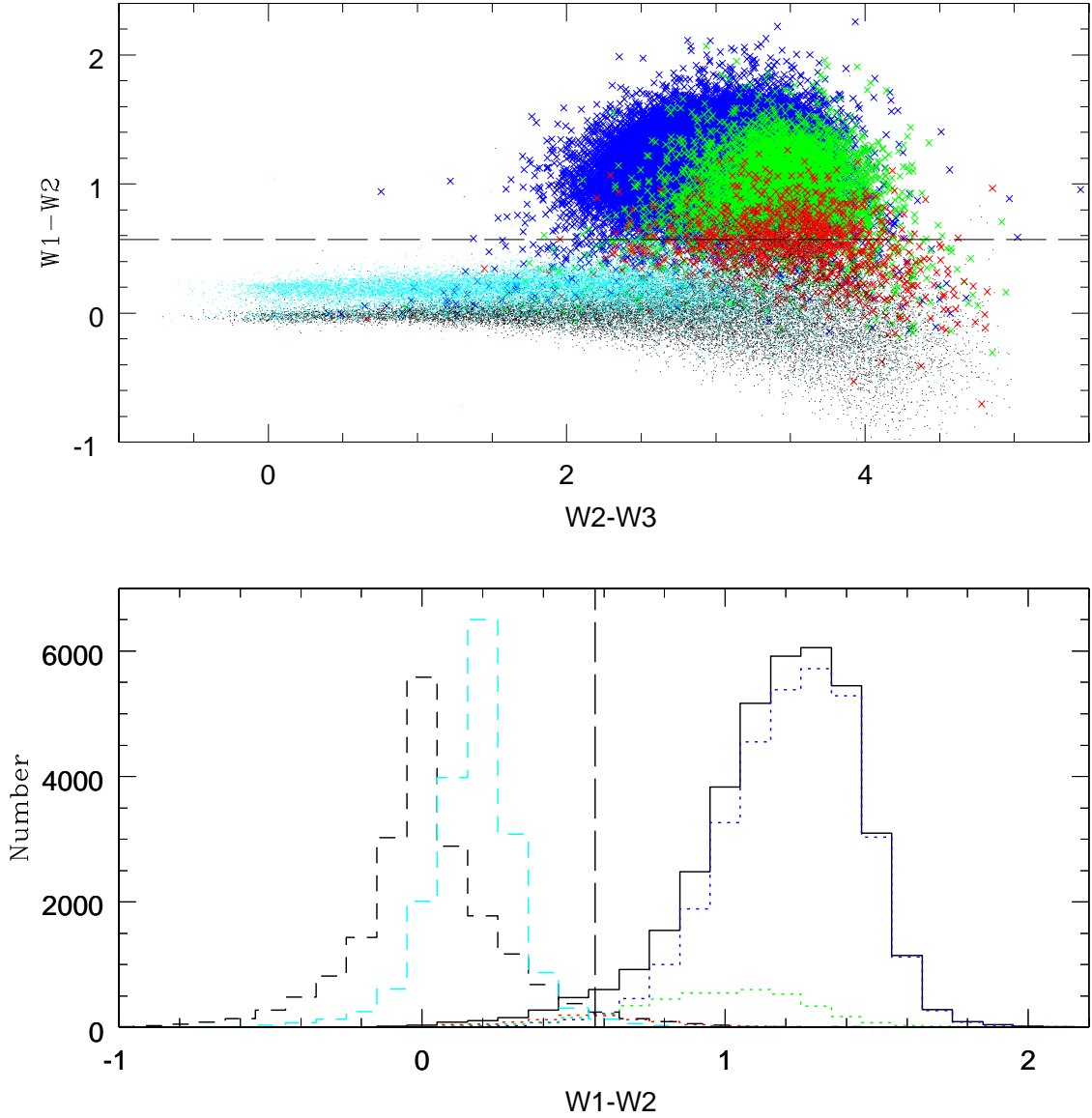}
\caption{Upper panel: The distributions of SDSS-WISE quasars and stars in the $W1-W2$ vs. $W2-W3$ 
color-color diagram. The blue, green and red crosses denote the quasars with redshifts of
$z<2.2$, $2.2<z<3.5$ and $z>3.5$, respectively. The black and cyan dots denote the
SDSS identified normal stars and late-type stars.   Lower panel: The histograms of $W1-W2$ colors
of SDSS-WISE quasars and stars. The black and cyan dashed lines denote the 
normal stars and late-type stars, while the blue, green and red dotted lines denote
quasars with redshifts of $z<2.2$, $2.2<z<3.5$ and $z>3.5$, respectively. The black
solid line marks the distribution of all quasars. The dashed lines in these
two panel indicate our proposed
 quasar selection criterion, $W1-W2>0.57$.
}
\end{figure}

\begin{figure}
\plotone{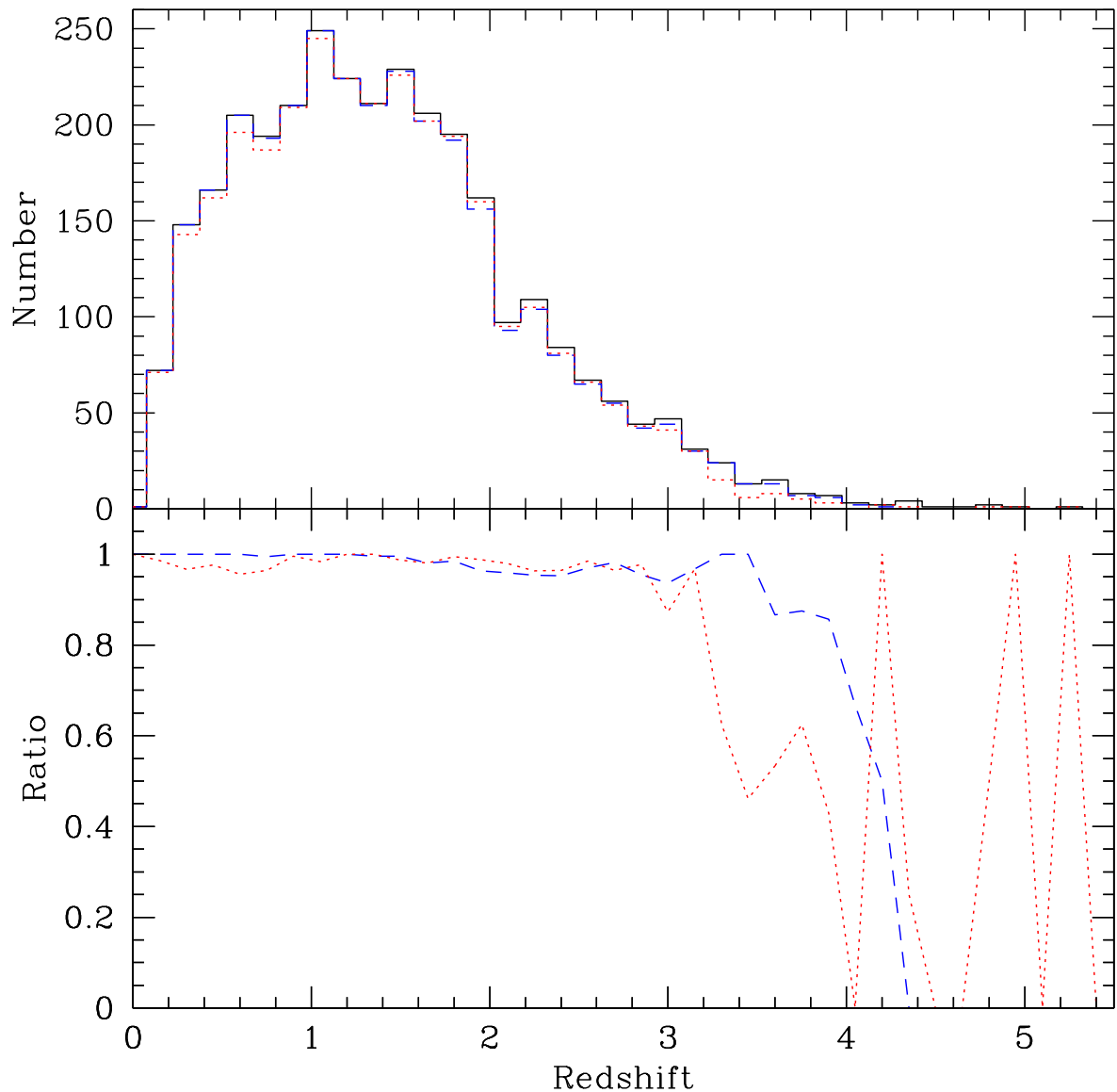}
\caption{Upper panel: Redshift distributions of FIRST radio-detected SDSS-WISE quasars 
(black solid line) and those selected by the criterion $z-W1>0.66(g-z)+2.01$ (blue dashed 
line) and by the criterion  $W1-W2>0.57$ (red dotted line). Lower panel: The completeness
ratio of radio quasars selected by $z-W1>0.66(g-z)+2.01$ (blue dashed 
line) and by criterion  $W1-W2>0.57$ (red dotted line) At different redshifts.
The ’noise’ behavior of the red dotted line at $z>3.5$ is due to the small number statistics.
}
\end{figure}

\begin{figure}
\plotone{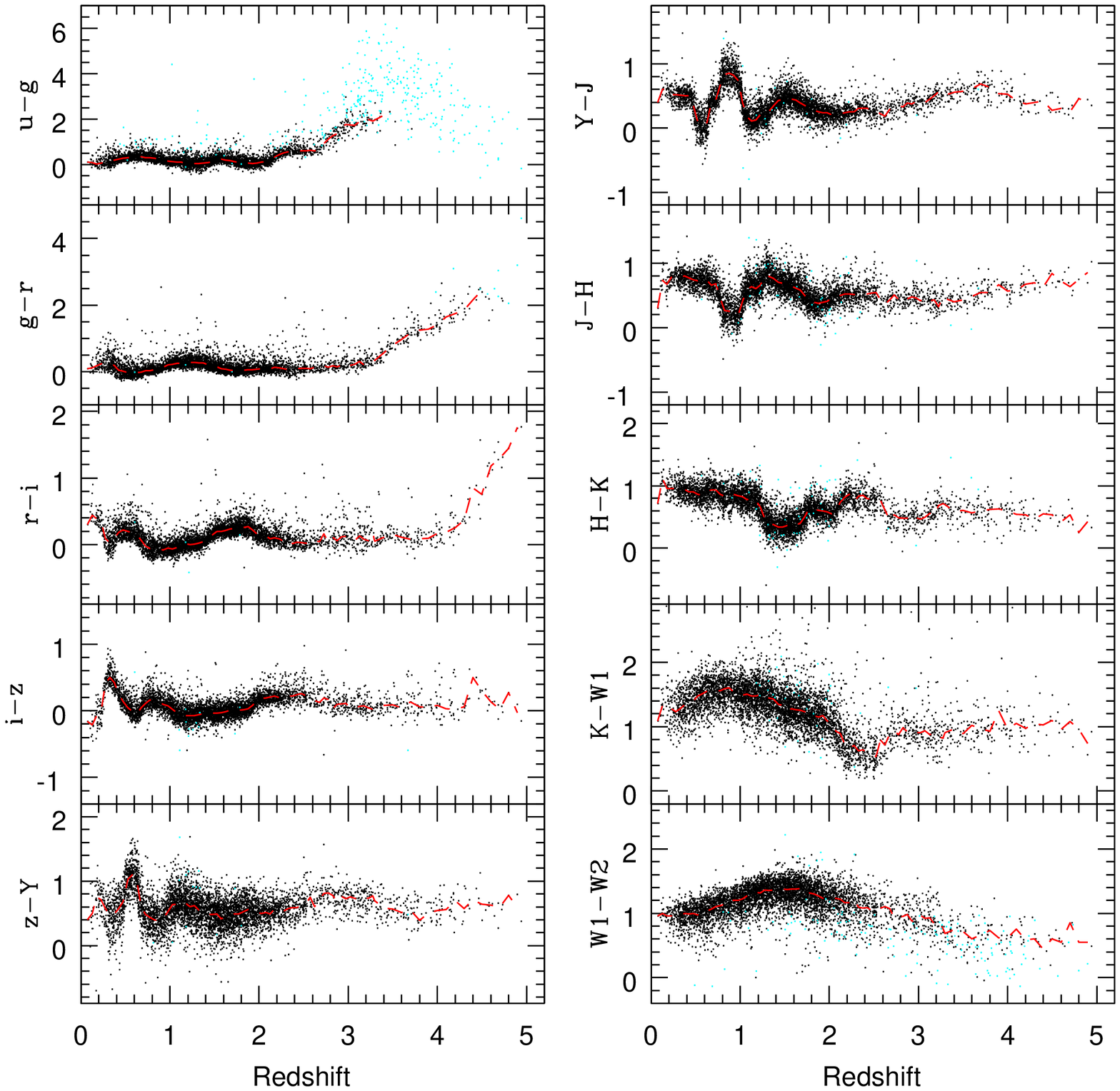}
\caption{Color-redshift relations of SDSS-UKIDSS-WISE quasars. The black dots
denote the quasars with magnitude uncertainties smaller than 0.2mag, and the cyan dots denote 
other quasars with
magnitude uncertainties larger than 0.2mag. The dashed lines represent the median color-redshift relations obtained from the quasars denoted as
black dots.
}
\end{figure}

\begin{figure}
\plotone{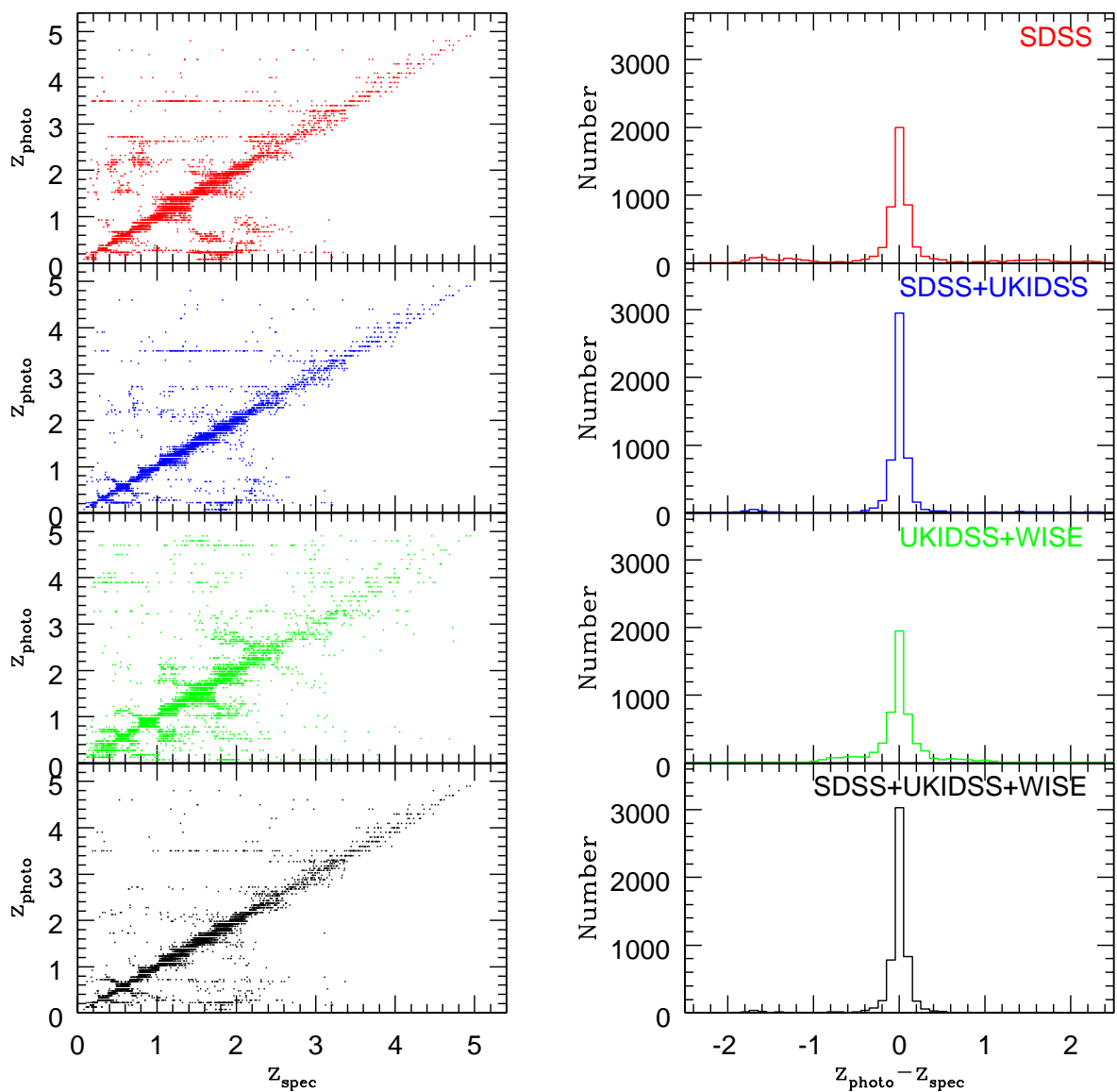}
\caption{Comparisons of photometric redshifts with spectroscopic redshifts (left panels)
and the distributions of the differences between them (right panels). The panels from top to bottom 
correspond to the cases 
using the SDSS, SDSS plus UKIDSS, UKIDSS plus WISE W1,W2, and SDSS plus UKIDSS plus
WISE W1,W2 photometric data, respectively.
}
\end{figure}

\end{document}